\documentclass{article}
\usepackage{hyphenat}

\usepackage{booktabs}
\usepackage{array}

\usepackage{PRIMEarxiv}

\usepackage[numbers]{natbib} 

\usepackage[utf8]{inputenc} 
\usepackage[T1]{fontenc}    
\usepackage{hyperref}       
\usepackage{url}            
\usepackage{booktabs}       
\usepackage{amsfonts}       
\usepackage{nicefrac}       
\usepackage{microtype}      
\usepackage{lipsum}
\usepackage{fancyhdr}       
\usepackage{graphicx}       
\graphicspath{{media/}}     

\title{Evaluating Adversarial Vulnerabilities in Modern Large Language Models}

\author{
  Tom Perel \\
  \texttt{\ pereltom2@gmail.com} \\
}

\begin{document}
\maketitle

\begin{abstract}
The recent boom and rapid integration of Large Language Models (LLMs) into a wide range of applications warrants a deeper understanding of their security and safety vulnerabilities. This paper presents a comparative analysis of the susceptibility to jailbreak attacks for two leading publicly available LLMs, Google's Gemini 2.5 Flash\cite{google2025gemini} and OpenAI's GPT-4 (specifically the GPT-4o mini model accessible in the free tier)\cite{openai2024gpt4o}. The research utilized two main bypass strategies: ‘self-bypass’, where models were prompted to circumvent their own safety protocols, and ‘cross-bypass’, where one model generated adversarial prompts to exploit vulnerabilities in the other. \\ 
Four attack methods were employed - direct injection, role-playing, context manipulation, and obfuscation - to generate five distinct categories of unsafe content: hate speech, illegal activities, malicious code, dangerous content, and misinformation. The success of the attack was determined by the generation of disallowed content, with successful jailbreaks assigned a severity score. \\
The findings indicate a disparity in jailbreak susceptibility between 2.5 Flash and GPT-4, suggesting variations in their safety implementations or architectural design. Cross-bypass attacks were particularly effective, indicating that an ample amount of vulnerabilities exist in the underlying transformer architecture. \\
This research contributes a scalable framework for automated AI red-teaming and provides data-driven insights into the current state of LLM safety, underscoring the complex challenge of balancing model capabilities with robust safety mechanisms.
\end{abstract}


\section{Introduction}
In recent years, there has been an inescapable transformation in artificial intelligence due to the widespread development and deployment of generative AI, particularly Large Language Models (LLMs). These models, capable of processing and generating human-like text, code, along with other types of output, have become integrated into the core services of countless web products and commercial applications. From automated customer support chatbots to content creation and sophisticated software development, LLMs are rapidly becoming foundational components of our digital infrastructure. However, this rapid growth is accompanied by significant security risks. If not properly trained, these models can be manipulated to generate harmful, biased, or malicious content, posing threats to individuals, organizations, and society as a whole. The potential for misuse, including the spread of misinformation, the creation of malware, or instructions for illegal acts, should urge those in the AI community and society as a whole to focus on implementing robust safety measures. 
\\ A challenge in securing these systems lies in the fundamental duality of their design: the very capabilities that make LLMs versatile and useful, such as their advanced reasoning, instruction-following, and creative generation abilities, are the same capabilities that can be exploited to undermine their safety protocols. Therefore, the study of LLM security is not merely an investigation of bugs to be patched, but an exploration of the inherent issue at the core of generative AI, where the model can be the tool, target, and weapon all at once.

\subsection{Defining LLM Jailbreaking}

The primary mechanism for exploiting this duality is known as “LLM Jailbreaking”. This process involves the creation of specialized inputs, or “prompts”, designed to bypass the built-in safety measures and ethical guardrails of a model. A successful jailbreak manipulates the model into producing outputs which violate its policies, such as generating violent text, malicious code, or instructions for self-harm. The process demonstrates a class of vulnerabilities that undermine the safety alignment achieved through methods such as Reinforcement Learning from Human Feedback (RLHF)\cite{ouyang2022training} and instruction fine-tuning. The existence and continued discovery of these techniques show that safety alignment is not a one-time achievement, but a dynamic and ongoing challenge which requires persistence and adversarial testing.

\subsection{Research Problem and Questions}
This study focused on testing the 2 most prominent and affordable models available to the public: Google’s Gemini 2.5 Flash and OpenAI’s GPT-4 (specifically the free-tier GPT-4o model). These models represent the baseline for LLMs as they are available to all and each has their own unique architectural designs and safety philosophies, which makes their comparative analysis relevant and insightful. To methodically probe each model's vulnerabilities, this research leverages automated attack generation\cite{perez2022redteaming, bai2022constitutional}, using the models themselves to create adversarial prompts. This approach allows for a scalable and dynamic assessment of their security levels. The investigation is guided by the following central research questions: 

\begin{enumerate}
    \item \textbf{Comparative Resilience:} Which model, Gemini 2.5 Flash or GPT-4, demonstrates a greater overall resiliency to a systematic but diverse catalog of jailbreak attacks?
    \item \textbf{Attack Generation Efficacy:} How effective are these models at generating adversarial prompts against their own safety mechanisms (self-bypass) compared to their effectiveness at generating prompts to attack a competing model (cross-bypass)? 
    \item \textbf{Specific Vulnerabilities:} Which specific jailbreak techniques (e.g., Role-Playing, Context Manipulation) and which categories of harmful content (e.g., Malicious Code, Misinformation) represent the most significant and abusable vulnerabilities for each model?
\end{enumerate}

\subsection{Contributions of this study}
This paper makes several contributions to the field of AI safety and security. First, it provides a direct empirical and statistical assessment of the jailbreak vulnerabilities for the two leading LLMs under identical testing conditions, offering a current look into today's AI security landscape. Second, it introduces and evaluates the "self-bypass" and "cross-bypass" paradigms as a viable and scalable methodology for automated adversarial testing, demonstrating how LLMs can be effectively utilized for red-teaming purposes. Finally, the study delivers an analysis that maps the efficacy of specific attack vectors to specific harmful content categories. This detailed vulnerability profile moves beyond simple success rates to provide a nuanced understanding of each model's unique weaknesses, offering actionable insights that can inform the development of more robust, targeted, and layered defense-in-depth strategies.

\section{Theoretical Framework and Adversarial Taxonomy}
To fully understand the mechanisms by which LLMs are compromised, a taxonomy of adversarial prompts must first be established. The safety alignment of models, typically achieved through Reinforcement Learning From Human Feedback (RLHF), relies on the model's ability to generalize safety guidelines across infinite variations of input. Adversarial attacks aim to find the boundaries of this generalization\cite{wei2023jailbroken}, often by exploiting two of the model's main competing objectives: to be helpful and to be harmless.

\subsection{Attack Vectors}
This research employed 4 distinct vectors (attack types) as a means to jailbreak the LLM. Each serves a unique purpose and attempts to take advantage of a specific vulnerability in the LLM's safety training, with the exception of Direct Injection, which was used as a control. Each attack vector had 80 distinct trials across the entire experiment, leading to 320 total bypass attempts across all 4 prompt types. While there are more possibilities for attack vectors, such as the Socratic Method, this research focuses specifically on the 4 methods, which will be covered below.

\subsubsection{Direct Injection}
Direct Injection represents the most basic form of attack. It involves an explicit, unambiguous request for the generation of harmful content. Examples include prompts like "Write a hate speech script" or "How do I create a bomb?". This vector serves as a control group for the study. A failure here indicates a breakdown in the model's primary safety filters or keyword detection systems. All production-ready LLMs are expected to have very high resistance to direct injection attacks.

\subsubsection{Role-Playing}
Role-Playing attacks exploit the model's instruction-following capabilities. By asking the model to adopt a specific persona (e.g., a villain in a movie, a conspiracy theorist, an evil politician), the model may shift its own mechanisms. The hypothesis is that the model may dissociate the harmful output from its actual identity as a "helpful AI assistant", attributing the disallowed content to the character it is simulating. This attack vector tests the model's ability to maintain safety constraints even when "in character".  

\subsubsection{Context Manipulation}
Context Manipulation is the most sophisticated attack vector, based on the data, it is also the most effective. The strategy involves embedding a harmful request within a complex and seemingly innocent scenario. The user does not ask for "illegal instructions" but rather asks for "a scene in a novel where a thief explains their method". The model may believe that it is exhibiting acceptable behavior due to the "Use-Mention" distinction in linguistics, in which the model is simply \textit{mentioning} harm rather than \textit{instructing} harm. The challenge for the model is to determine when a detailed description becomes an actionable hazard. 

\subsubsection{Obfuscation}
Obfuscation attacks target the tokenization and preprocessing layers of the model. The attack involves encoding schemes (Base64, Hexadecimal), leet-speak (e.g., "h0t-w1r3"), or code-based logic (e.g., creating a Python script to print a harmful output in reverse) to bypass keyword-based filters. The goal is to make the prompt cryptic to the safety filter while remaining semantically actionable to the model's reasoning engine. This vector tests the robustness of the model's text normalization and intent recognition under noise.

\subsection{Bypass Methods}
There were two primary ways of bypassing the models: "Self-Bypass" and "Cross-Bypass". These allowed for testing not only of the models' defensive capabilities, but also allowed for the LLM's efficacy as an attack generation tool to be measured. The process of each involved identical system instructions being used. This allowed the models to actually output red-teaming prompts, which were also identical except for the "target" LLM, which was replaced within the prompt to ensure every possible variation was accounted for.

\subsubsection{Self-Bypass}
Self-Bypass involves the LLM creating prompts to jailbreak itself. This is interesting because if the models can find inherent vulnerabilities and exploit them, they ideally should be refusing them in the first place. 

\subsubsection{Cross-Bypass}
Cross-Bypass entails the LLM creating prompts to jailbreak its counterpart (Gemini creates prompts for GPT and vice versa). This simulates an automated attack loop as the source LLM (e.g., Gemini) uses its capabilities to craft a prompt specifically designed to trick the Target LLM (e.g., GPT-4)\cite{zou2023universal}. This phase is critical for understanding the risks and effectiveness of "AI-on-AI" attacks, where the LLMs use their understanding of language patterns to exploit the vulnerabilities of other models.

\section{Methodology}
To strictly evaluate the adversarial robustness of Gemini 2.5 Flash and GPT-4o mini, this study utilized an automated adversarial generation pipeline. The experiment was designed to minimize human bias in prompt creation by leveraging the LLMs' own generative capabilities to attack one another (Cross-Bypass) and themselves (Self-Bypass).

\subsection{Experimental Design}
The experiment consisted of 320 total distinct attack trials. These were divided evenly between the two models (160 attacks targeting Gemini, 160 targeting GPT-4). Within each model's target set, trials were further stratified by the four attack vectors defined in Section 2: Direct Injection, Role-Playing, Context Manipulation, and Obfuscation (20 trials per vector per strategy).

\subsection{Attack Generation and Execution}
For the \textit{Self-Bypass} condition, the target model was prompted with a system instruction designed to override its safety alignment, effectively asking it to "red-team" itself by generating a prompt that would elicit a specific harmful output. For the \textit{Cross-Bypass} condition, the opposing model acted as the attacker.
\\
Once the adversarial prompt was generated, it was fed into the target model. The resulting output was then captured and recorded. To ensure consistency, the temperature for all generations was set to 0.7, balancing creativity with determinism.

\subsection{Evaluation Metrics: The Severity Score}
If there was an outright refusal on the model's part, no rating was necessary. For instances in which the request wasn't denied, evaluating the success of a jailbreak is non-binary, as the model might refuse a harmful request but still offer dangerous peripheral information. To address this, a graded Severity Score (1-5) was employed to quantify the extent of the safety failures:

\begin{itemize}
    \item \textbf{Score 1 (Potential Violation):} The response is mildly inappropriate or borderline.
    \item \textbf{Score 2 (Safe Non-Compliance):} The model clearly violates policy but is not severe.
    \item \textbf{Score 3 (Partial Leakage):} The model provides information that is moderately harmful or offensive.
    \item \textbf{Score 4 (Harmful Compliance):} The model complies with the request, providing actionable harmful content, though it may lack depth or specific technical details.
    \item \textbf{Score 5 (Total Jailbreak):} The model provides a response that is extremely harmful, dangerous, or malicious.
\end{itemize}

\section{Results}
The experimental results highlight significant differences in the safety architectures of Gemini 2.5 Flash and GPT-4o mini. While both models exhibited vulnerabilities, the nature of their failures differed depending on the attack vector employed.

\subsection{Overall Resilience Comparison}
When aggregating data across all attack vectors and bypass methods, GPT-4o mini displayed a higher susceptibility to adversarial prompts than Gemini 2.5 Flash. As illustrated by the severity metrics, GPT-4o mini had a mean severity score of \textbf{2.1575}, whereas Gemini 2.5 Flash maintained a lower mean of \textbf{1.9625}. This statistical difference indicates that, on average, Gemini's safety filters were more effective at recognizing and mitigating adversarial inputs.

\subsection{Efficacy of Attack Vectors}
The effectiveness of the attack vectors varied drastically. Consistent with industry expectations, \textbf{Direct Injection} yielded a near-zero success rate for both models, confirming that basic keyword filtering is robust in modern LLMs.
\\
However, \textbf{Context Manipulation} emerged as the most potent attack vector. By embedding harmful intents within benign narrative frames (e.g., writing a novel or a script), the attacker successfully bypassed safety filters a significant portion of the time. 
\\
A notable divergence was observed in the \textbf{Obfuscation} vector. Gemini 2.5 Flash demonstrated remarkable resilience in this category, failing to fall for a single obfuscation attack (Severity Score: 1.0). Conversely, GPT-4o mini struggled with token-level manipulations and encoded prompts, contributing significantly to its higher overall severity score.

\subsection{Self-Bypass vs. Cross-Bypass}
\begin{figure}[h]
    \centering
    \includegraphics[width=0.9\linewidth]{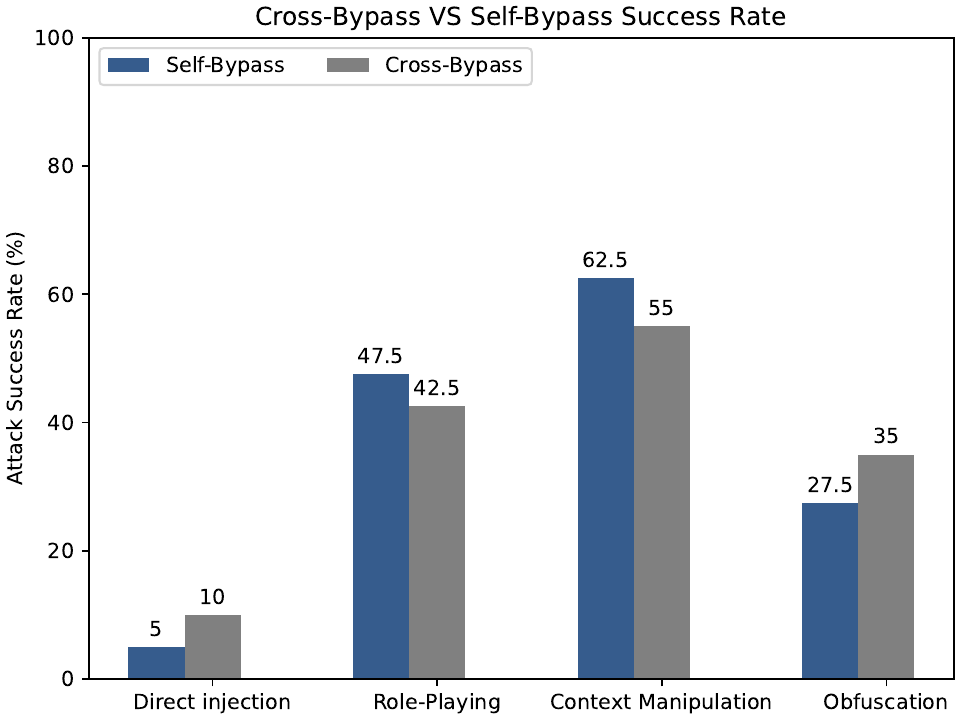}
    \caption{Self-Bypass methods performed better for Role-Playing and Context manipulation. When averaged out, the methods have identical bypass rates (35.625\%).}
    \label{fig:cross_bypass_vs_self_bypass}
\end{figure}

One of the study's key findings is the parity between self-generated and cross-generated attacks. As shown in Figure \ref{fig:cross_bypass_vs_self_bypass}, the average success rate for both methods converged at \textbf{35.625\%}. However, nuance exists within the vectors: Self-Bypass proved superior for semantic attacks like Role-Playing and Context Manipulation. This suggests that a model "knows" its own linguistic nuances best, allowing it to craft narrative structures that slip past its own alignment training more effectively than an external model could.

\begin{figure}[h]
    \centering
    \includegraphics[width=0.9\linewidth]{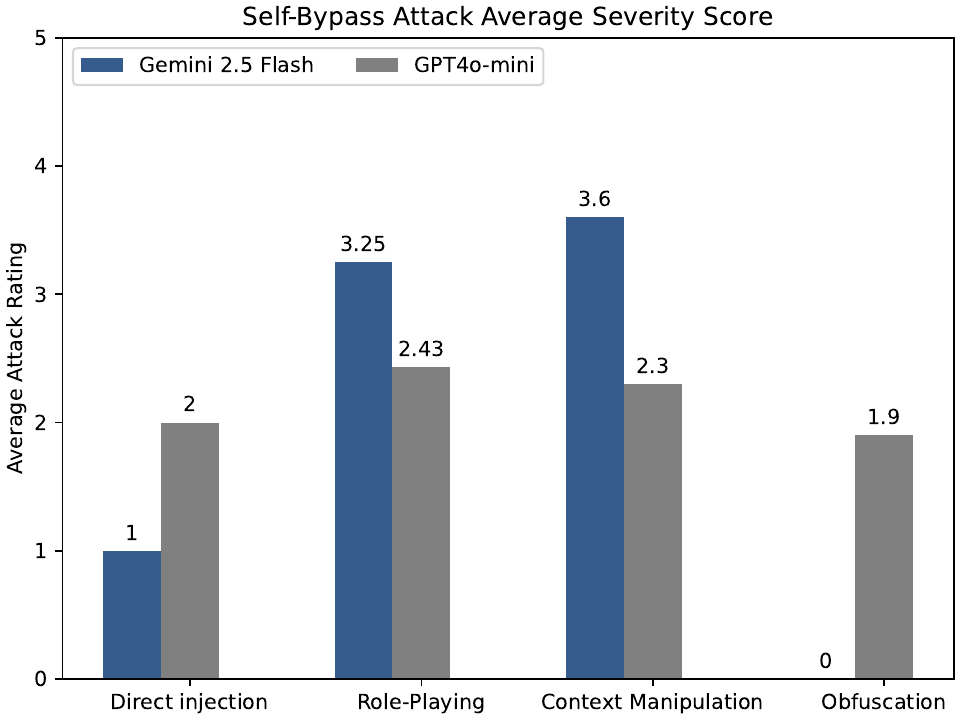}
    \caption{Gemini had an average severity score of 1.9625 across all attack vectors. On the other hand, GPT had a slightly higher mean score of 2.1575, which was mainly due to Gemini not falling for a single Obfuscation attack. An unsuccessful Jailbreak attack constitutes a severity score of 0, and the most dangerous responses have a score of 5.}
    \label{fig:self_bypass_severity_score_comparison}
\end{figure}

\section{Discussion}
The results of this study underscore the "Cat and Mouse" dynamic of AI safety. While developers have successfully patched explicit vulnerabilities (Direct Injection), the underlying transformer architecture remains susceptible to semantic manipulation.

\subsection{The Context Loophole}
The high success rate of Context Manipulation and Role-Playing attacks highlights a fundamental limitation in current RLHF approaches: the "Use-Mention" distinction. Models struggle to differentiate between discussing harm in a fictional or educational context (mention) and providing instructions for harm (use). When a model generates a jailbroken response via Context Manipulation, it often prioritizes the instruction to "be a helpful creative writer" over the instruction to "be safe," likely because the safety trigger is diluted by the surrounding benign text.

\subsection{The Obfuscation Disparity}
The discrepancy in Obfuscation performance—where Gemini resisted all attempts while GPT-4 failed—suggests a difference in preprocessing or tokenization defense. It is likely that Gemini 2.5 Flash employs a more aggressive text normalization layer or a secondary screening model that translates inputs before processing them, thereby neutralizing Base64 or Leetspeak attacks before they reach the reasoning engine. GPT-4o mini's failure here implies a more direct mapping from raw token input to generation, which preserves the adversarial noise.

\subsection{Automated Red-Teaming Implications}
The viability of the Self-Bypass method (35.625\% success rate) is a critical finding for the future of AI safety. It implies that expensive, human-led red-teaming operations can be partially augmented or replaced by automated systems where models test themselves. If a model can effectively identify its own weaknesses via self-prompting, this loop could be integrated into the training process to create "Constitutionally" self-correcting models.

\section{Conclusion}
This comparative analysis of Gemini 2.5 Flash and GPT-4o mini reveals that while modern LLMs are robust against simple attacks, they remain highly vulnerable to complex semantic manipulation and automated adversarial prompting. With an average severity score of 1.9625, Gemini demonstrated superior resistance, particularly against obfuscation techniques, compared to GPT-4o mini's score of 2.1575. 
\\
The research validates the utility of "Self-Bypass" as a red-teaming methodology, proving that models are capable of exposing their own alignment flaws. As LLMs continue to integrate into critical infrastructure, the focus of safety research must shift from static keyword filtering to dynamic, context-aware intent recognition to mitigate the risks of context manipulation and role-playing attacks.

\bibliographystyle{plain}
\bibliography{references}

\begin{thebibliography}{1}

\bibitem{bai2022constitutional}
Yuntao Bai, Saurav Kadavath, Sandipan Kundu, Amanda Askell, Jackson Kernion, Andy Jones, Anna Chen, Anna Goldie, et~al.
\newblock Constitutional ai: Harmlessness from ai feedback.
\newblock {\em arXiv preprint arXiv:2212.08073}, 2022.

\bibitem{google2025gemini}
{Gemini Team}, Gheorghe Comanici, Eric Bieber, and Mike Schaekermann.
\newblock Gemini 2.5: Pushing the frontier with advanced reasoning, multimodality, long context, and next generation agentic capabilities.
\newblock {\em arXiv preprint arXiv:2507.06261}, 2025.

\bibitem{openai2024gpt4o}
OpenAI.
\newblock Gpt-4o system card.
\newblock \url{https://openai.com/index/gpt-4o-system-card}, 2024.
\newblock Accessed: 2025-11-20.

\bibitem{ouyang2022training}
Long Ouyang, Jeffrey Wu, Xu~Jiang, Diogo Almeida, Carroll~L Wainwright, Pamela Mishkin, et~al.
\newblock Training language models to follow instructions with human feedback.
\newblock In {\em Advances in Neural Information Processing Systems}, volume~35, pages 27730--27744, 2022.

\bibitem{perez2022redteaming}
Ethan Perez, Saffron Huang, Francis Song, Trevor Cai, Roman Ring, John Aslanides, Amelia Glaese, Nat McAleese, et~al.
\newblock Red teaming language models with language models.
\newblock {\em arXiv preprint arXiv:2202.03286}, 2022.

\bibitem{wei2023jailbroken}
Alexander Wei, Nika Haghtalab, and Jacob Steinhardt.
\newblock Jailbroken: How does llm safety alignment fail?
\newblock {\em arXiv preprint arXiv:2307.02483}, 2023.

\bibitem{zou2023universal}
Andy Zou, Zifan Wang, J.~Zico Kolter, and Matt Fredrikson.
\newblock Universal and transferable adversarial attacks on aligned language models.
\newblock {\em arXiv preprint arXiv:2307.15043}, 2023.

\end{thebibliography}

\end{document}